\documentclass[pra,showpacs,showkeys,amsfonts,amsmath,twocolumn]{revtex4}
\usepackage{bm}
\usepackage{graphicx}
\RequirePackage{mathptm}

\numberwithin{equation}{section}

\newcommand{\ip}[2]{\langle \,{#1},\,{#2}\,\rangle}
\newcommand{\W}[4]{\begin{cases}
#1 ,&#2\\[2mm]
#3 ,&#4
\end{cases}}
\newcommand{\ro}{\varrho}

\newcommand{\la}{\lambda}
\newcommand{\las}[1]{\lambda_{#1}}

\newcommand{\al}{\alpha}

\newcommand{\ee}{\varepsilon}

\newcommand{\te}{\theta}
\newcommand{\vf}{\varphi}

\newcommand{\I}{\openone}

\newcommand{\fala}[1]{\widetilde{#1}}
\newcommand{\ket}[1]{|{#1}\rangle}
\newcommand{\bra}[1]{\langle {#1} |}

\newcommand{\C}{\mathbb C}
\newcommand{\R}{\mathbb R}

\newcommand{\PP}{\mathbb P}

\newcommand{\cA}{{\mathcal A}}
\newcommand{\cB}{{\mathcal B}}

\newcommand{\tr}{\mathrm{tr}\,}
\newcommand{\ptr}[1]{\mathrm{tr}_{#1}}

\newcommand{\mr}[1]{\mathrm{#1}}

\newcommand{\DS}{\displaystyle}

\begin{document}
\title{Qutrit geometric discord}
\author{  Lech Jak{\'o}bczyk
\footnote{ ljak@ift.uni.wroc.pl}, Andrzej Frydryszak \footnote{ amfry@ift.uni.wroc.pl} and Piotr
{\L}ugiewicz }
 \affiliation{Institute of Theoretical Physics\\ University of
Wroc{\l}aw\\
Plac Maxa Borna 9, 50-204 Wroc{\l}aw, Poland}
\begin{abstract}
Properties of the trace norm geometric discord of
the system of two qutrits are studied. The geometric discord of qutrit Bell states,  Werner states and bound
entangled states is computed.
\end{abstract}
\maketitle
\section{Introduction}
Quantum correlations in finite-dimensional composite systems have
been studied intensively in last decades, originally with  focus on  the simplest  systems i.e. qubit systems.
Although one can consider in this respect a generic d-level system, so called qudit, but progress
in studying quantum correlations for qudits  is limited, due to the level of complication.
Therefore,  three-level systems (qutrits) are intensively studied presently. They are
interesting for many reasons. First of all, such systems model realistic three-level atoms
in which the interference between different radiative transitions is
possible, resulting in  new kinds of coherence \cite{FS}. Quantum
dynamics of collective systems of such atoms significantly differ
from a  dynamics of two-level atoms (see e.g. \cite{AP, EKM, DJ}).
On the other hand, the theory of quantum correlations between the
pairs of such atoms is much more complex than in the case of qubits.
Even description of the set of states of a single qutrit is much more
complicated then that for qubit states \cite{GSS}. Moreover, there is no simple
necessary and sufficient condition probing entanglement of qutrits.
The Peres-Horodecki
separablity criterion \cite{Pe, HHH1} is not sufficient for two qutrit system,
it only shows that the states that are not positive after partial transposition
(NPPT states) are entangled. It turns out that all entangled states can be divided into two
classes: free entangled states that can be distilled using local operations and classical communication (LOCC);
bound entangled states for which no LOCC strategy can be used to extract pure state entanglement \cite{HHH2}.
Since many effects in quantum information depend on maximally entangled pure states, only distillable states can directly be used for quantum communication.
\par
Recently, more general properties of quantum correlations, which go beyond quantum entanglement, have attracted a lot of interest. They arise from the observation that for pure separable states, there exists von Neumann measurements on a part of composite system that do not disturb the state, whereas nonseparable states are always disturbed by such local measurements. Extension of this feature to the mixed states, gives rise to the notion of quantum discord \cite{OZ, HV, MBC}. For pure states notion of quantum discord coincides with entanglement, but in the case of mixed states discord and entanglement differ significantly. For example, almost all quantum states have non-vanishing discord and there exist discordant separable mixed states \cite{FA}.
\par
To evaluate quantum discord at a given state, one can use its geometric measure instead of the original measure proposed in \cite{OZ}. Such geometric measure of quantum discord is given by a minimal distance of a given state $\ro$  from the set of states  $\PP_{\cA}(\ro)$, obtained after any local measurement $\PP_{\cA}$ on a part $\cA$. The proper choice of a distance measure is crucial. Presently there are three of them in use. The measure proposed in \cite{DVB} uses a Hilbert-Schmidt norm to define a distance in the set of states. This choice has a technical advantage: the minimization process can be realized analytically for arbitrary two-qubit states. However this measure has some unwanted properties. The most important problem is that it may  increase under local operations performed on the unmeasured system \cite{P,TGV, J}. It can be cured by the use of the Schatten 1-norm (trace norm) instead, however such defined measure of quantum discord is more difficult to compute \cite{Paula}. By now, the explicit formula for it is known only in the case of Bell-diagonal states or two-qubit X-states \cite{Paula, Cic}. The third measure used  for studying the geometric quantum discord is based on the Bures distance \cite{SpOr}. It has nice property that for pure states it is strictly equal to the geometric measure of entanglement.
\par
In this paper, we extend the analysis of trace norm based geometric discord $D_{1}$  to the system of two qutrits. The results known for the two qutrit system are related to the geometric discord based on the Hilbert-Schmidt norm and give information only about two qutrit Werner  states \cite{Y} and about upper and lower bounds of such discord in the case of bound entangled states
\cite{RP}. Firstly we compute the form of $D_{1}$ for special class of states with maximally mixed marginals and diagonal correlation matrix. We find that for pure Bell states and qutrit Werner states, the distance of a state $\ro$ to states $\PP_{\cA}(\ro)$ is constant, and  one does not need to minimize over all local measurements to  compute quantum discord. The value of this distance for the Bell state we use to normalize $D_{1}$ in such a way that for any state $\ro$
\begin{equation*}
0\leq D_{1}(\ro)\leq 1
\end{equation*}
and $D_{1}(\ro)=1$ for  maximally entangled state $\ro$. Then, the
normalized quantum discord is computed for a class of qutrit Werner
states $\ro_{W}$ and we obtain the result that
$D_{1}(\ro_{W})$ is equal to the mixing parameter $p$ (Sect. IV.B).
For other families of states with maximally mixed marginals, a
minimization over all measurements is necessary. This makes the
problem of analytic evaluation of qutrit discord extremely
difficult. Fortunately, for examples considered in this work,
numerical analysis shows that the minimal distance between $\ro$ and
$\PP_{\cA}(\ro)$ is achieved for projective measurement given by
standard orthonormal basis in $\C^{3}$. In this way we can compute trace
norm geometric discord for  two-parameter family of mixed
entangled states  (Sect. V.A) and one-parameter family containing
bound entangled states (Sect. V.B). In particular, we obtain first known
analytic formula giving trace norm quantum discord of bound
entangled states.
\section{Qudit state parametrization}
\subsection{One - qudit parametrization}
Let us start our analysis with the general $d$-level quantum
system (qudit). To describe the states of qudit, it is convenient to
use as a basis in a set of $d\times d$ matrices the hermitian
generators of $\mr{su(d)}$ algebra and the identity matrix $\I_{d}$.
Let $\las{1},\ldots,\las{d^{2}-1}$  be the generators of
$\mr{su(d)}$ algebra.   The matrices $\las{j}$ satisfy
\begin{equation*}
\tr\, \las{j}=0,\quad \tr\, (\las{j}\las{k})=2\,\delta_{jk},\;
j,k=1,\ldots,d^{2}-1
\end{equation*}
and
\begin{equation}
\las{j}\las{k}=\frac{2}{d}\,\delta_{jk}\,\I_{d}
+\sum\limits_{l}\,(\hat{d}_{jkl}+i\,\hat{f}_{jkl})\,\las{l}\label{lajlak}
\end{equation}
where the structure constants $\hat{d}_{jkl}$ and $\hat{f}_{jkl}$
are given by
\begin{equation}
\hat{d}_{jkl}=\frac{1}{4}\,\tr\,([\las{j},\las{k}]_{+}\,\las{l})\label{d}
\end{equation}
and
\begin{equation}
\hat{f}_{jkl}=\frac{1}{4i}\,\tr\,([\las{j},\las{k}]\,\las{l})\label{f}
\end{equation}
Using the structure constants (\ref{d}) and (\ref{f}) one can
introduce the following "star" and "wedge" products in a real linear
space  $\R^{d^{2}-1}$. For $n,\, m\in \R^{d^{2}-1}$ we define
\begin{equation}
(n\star
m)_{j}=\sqrt{\frac{d(d-1)}{2}}\,\frac{1}{d-2}\,\sum\limits_{k,l}\,\hat{d}_{jkl}n_{k}m_{l}
\end{equation}
and
\begin{equation}
(n\wedge
m)_{j}=\sqrt{\frac{d(d-1)}{2}}\,\frac{1}{d-2}\,\sum\limits_{k,l}\,\hat{f}_{jkl}n_{k}m_{l}
\end{equation}
Let $\la=(\las{1},\ldots,\las{d^{2}-1})$ and
\begin{equation}
\ip{n}{\la}=\sum\limits_{j}n_{j}\las{j}
\end{equation}
then taking into account (\ref{lajlak}), we obtain
\begin{equation}
\ip{n}{\la}\ip{m}{\la}=\frac{2}{d}\,\ip{n}{m}\I_{d}+\frac{1}{d^{\prime}}\,\ip{n\star
m}{\la}+\frac{i}{d^{\prime}}\,\ip{n\wedge m}{\la}, \label{product}
\end{equation}
where
\begin{equation*}
d^{\prime}=\sqrt{\frac{d(d-1)}{2}}\,\frac{1}{d-2}.
\end{equation*}
The set of states of $d$-level system can be parametrized as
follows (see e.g. \cite{BK})
\begin{equation}
\ro=\frac{1}{d}\,\left(\I_{d}+d^{\prime\prime}\,\ip{n}{\la}\right),\quad
n\in \R^{d^{2}-1},\label{state}
\end{equation}
where
\begin{equation*}
d^{\prime\prime}=\sqrt{\frac{d(d-1)}{2}}
\end{equation*}
and the components of the vector $n$ are
\begin{equation*}
n_{j}=\frac{d}{\sqrt{2d(d-1)}}\,\tr\,(\ro\,\la_{j}),\quad
j=1,\ldots,d^{2}-1.
\end{equation*}
The matrix (\ref{state}) is hermitian and has a unit trace. To describe a quantum state, the matrix $\ro$ have to be
positive-definite
and this condition is not easy to characterize in terms of the vector $n$. However the pure states given
by one-dimensional projectors
can be fully described. Using (\ref{product}), one can check that $\ro$ given by (\ref{state}) satisfies
$\ro^{2}=\ro$ if and only if
\begin{equation*}
\ip{n}{n}=1\quad\text{and}\quad n\star n=n.
\end{equation*}
\subsection{Two  - qudit parametrization}
Consider now two qudits $\cA$ and $\cB$. The state of a compound
system can be parametrized as follows
\begin{equation}
\begin{split}
\ro=\frac{1}{d^{2}}\bigg(&\I_{d}\otimes
\I_{d}+d^{\prime\prime}\,\ip{x}{\la}\otimes \I_{d}+\I_{d}\otimes
d^{\prime\prime}\,\ip{y}{\la}\\
&+\sum\limits_{j,k=1}^{d^{2}-1}T_{jk}\las{j}\otimes\las{k}\bigg)
\end{split}
\label{9state}
\end{equation}
with $x,\, y\in \R^{d^{2}-1}$. Notice that
\begin{equation*}
x_{j}=\frac{d}{\sqrt{2d(d-1)}}\,\tr\,(\ro\,\las{j}\otimes\I_{d}),\quad
y_{j}=\frac{d}{\sqrt{2d(d-1)}}\,\tr\,(\ro\,\I_{d}\otimes \las{j})
\end{equation*}
and
\begin{equation*}
T_{jk}=\frac{d^{2}}{4}\,\tr\,(\ro\las{j}\otimes\las{k}).
\end{equation*}
The parametrization (\ref{9state}) is chosen is such a way, that the marginals
$\ptr{\cA}\ro$ and $\ptr{\cB}\ro$ are given by the vectors $x$ and $y$ as in (\ref{state}).
\section{Trace-norm geometric  qudit discord}
When a bipartite system $\cA\cB$ is prepared in a state $\ro$ and we
perform local measurement on the subsystem $\cA$, almost all states
$\ro$ will be disturbed due to such measurement. The one-sided
geometric discord is defined as the minimal disturbance induced by
any projective measurement $\PP_{\cA}$ on subsystem $\cA$
\cite{DVB}. Here we choose a distance in the set of states given by
the trace norm, instead of Hilbert-Schmidt norm used in the
standard approach, and define (not normalized) measure of quantum
discord  as\cite{Paula}
\begin{equation}
\fala{D}_{1}(\ro)=\min\limits_{\PP_{\cA}}\,||\ro-\PP_{\cA}(\ro)||_{1},\label{trdisc}
\end{equation}
where
\begin{equation*}
||\sigma||_{1}=\tr\,|\sigma|.
\end{equation*}
In the case of qudits, local projective measurement $\PP_{\cA}$ is
given by the one-dimensional  projectors $P_{1},\, P_{2},\ldots,\,
P_{d}$ on $\C^{d}$, such that
\begin{equation*}
P_{1}+P_{2}+\cdots + P_{d}=\I_{d},\quad
P_{j}P_{k}=\delta_{jk}\,P_{k}
\end{equation*}
and $\PP_{\cA}=\PP\otimes \mr{id}$, where
\begin{equation*}
 \PP(\sigma)=P_{1}\,\sigma\, P_{1}+P_{2}\,\sigma\, P_{2}+\cdots +P_{d}\,\sigma\,P_{d}.
\end{equation*}
It is worth to stress that definition (\ref{trdisc}) is equivalent to the more common one, which is
given by the minimal distance of a given state to the set $\Omega_{0}$ of all states with zero discord. In the case of one-sided
quantum discord studied in this paper, the set $\Omega_{0}$ contains all "classical-quantum" states
\begin{equation*}
\ro_{\mr{cq}}=\sum\limits_{k=1}^{3}p_{k}\ket{\psi_{k}}\bra{\psi_{k}}\otimes \ro_{k}^{\cB},
\end{equation*}
where $\{\psi_{k}\}$ is any single-qutrit orthonormal basis, $\{\ro_{k}^{\cB}\}$ are any states of the subsystem $\cB$ and
$p_{k}\geq 0,\; \sum\limits_{k=1}^{3}p_{k}=1$.
\par
For the state (\ref{9state}) we have
\begin{equation*}
\begin{split}
\ro-\PP_{\cA}(\ro)=\frac{1}{d^{2}}\,&\big[\,d^{\prime\prime}\ip{x}{\la}-
\PP(d^{\prime\prime}\ip{x}{\la}))\otimes \I_{d}\\
&+\sum\limits_{j,k=1}^{d^{2}-1}T_{jk}\,(\las{j}-
\PP(\las{j}))\otimes \las{k}\,\big].
\end{split}
\end{equation*}
Since
\begin{equation*}
\PP(\las{j})=\sum\limits_{k=1}^{d^{2}-1}a_{jk}\,\las{k},\quad
a_{jk}=\frac{1}{2}\,\tr\,(\PP(\las{j})\las{k})
\end{equation*}
and the matrix $A=(a_{jk})$ is  real  and symmetric (in fact it is a projector operator \cite{ZCL}),
\begin{equation*}
\PP(\ip{m}{\la})=\ip{m}{A\la}=\ip{Am}{\la},\quad m\in \R^{d^{2}-1}.
\end{equation*}
So
\begin{equation}
\ro-\PP_{\cA}(\ro)=\frac{1}{d^{2}}\,\big[
d^{\prime\prime}\ip{Mx}{\la}\otimes
\I_{d}+\sum\limits_{j,k}T_{jk}\ip{Me_{j}}{\la}\otimes
\ip{e_{k}}{\la}\big] \label{dif}
\end{equation}
where $M=\I_{d^{2}-1}-A$ and $\{e_{j}\}_{j=1}^{d^{2}-1}$ is the
canonical basis in $\R^{d^{2}-1}$. Let $R(M)$ denotes the right hand
side of equation (\ref{dif}). Then  not normalized geometric quantum
discord of the state (\ref{9state}) equals
\begin{equation}
\fala{D}_{1}(\ro)=\min\limits_{M}||R(M)||_{1}=\min\limits_{M}\,\tr\,\sqrt{Q(M)}
\end{equation}
where $Q(M)=R(M)\,R(M)^{\ast}$ and the minimum is taken over all
matrices $M$ corresponding to a measurements on subsystem $\cA$.
\par
To simplify further computations, we consider first the states with
maximally mixed marginals i.e. such states $\ro$ that
\begin{equation}
\ptr{\cA}\ro=\frac{\I_{d}}{d},\quad \ptr{\cB}\ro=\frac{\I_{d}}{d}.\label{MMM}
\end{equation}
In the parametrization (\ref{9state}) this property corresponds to $x=y=0$.
We  also choose such states
for which the correlation matrix $T=(T_{jk})$ is diagonal. (Notice that contrary to the case of qubits ($d=2$),
the general state of two qudits  ($d>2$) satisfying (\ref{MMM}) is not locally equivalent to the state with diagonal $T$
and such defined class is only a subclass of all states with maximally mixed marginals.)  Let
\begin{equation*}
T=\mr{diag}\,(t_{1},\ldots,t_{d^{2}-1}),
\end{equation*}
then
\begin{equation}
R(M)=\frac{1}{d^{2}}\,\sum\limits_{j=1}^{d^{2}-1}t_{j}\,\ip{M\,e_{j}}{\la}\otimes
\ip{e_{j}}{\la},
\end{equation}
and using (\ref{product}), we obtain
\begin{equation*}
\begin{split}
Q(M)=\frac{1}{d^{4}}\,\bigg[&\,\frac{4}{d^{2}}\,\sum\limits_{j}\,t_{j}^{2}\,\ip{Me_{j}}{Me_{j}}\,
\I_{d}\otimes \I_{d}\\
&+\frac{2}{d\,d^{\prime}}\,\sum\limits_{j}\,t_{j}^{2}\,\ip{Me_{j}\ast Me_{j}}{\la}\otimes\I_{d}\\
&+\frac{2}{d\,d^{\prime}}\,\sum\limits_{j,k}t_{j}t_{k}\,\ip{Me_{j}}{Me_{k}}\I_{d}\otimes \ip{e_{j}\ast e_{k}}{\la}\\
&+\frac{1}{d^{\prime 2}}\,\sum\limits_{j,k}t_{j}t_{k}\,\ip{Me_{j}\ast Me_{k}}{\la}\otimes \ip{e_{j}\ast e_{k}}{\la}\\
&-\frac{1}{d^{\prime
2}}\sum\limits_{j,k}t_{j}t_{k}\,\ip{Me_{j}\wedge Me_{k}}{\la}\otimes
\ip{e_{j}\wedge e_{k}}{\la}\,\bigg]
\end{split}
\end{equation*}
\section{The case of qutrits}
Now we consider in more details the case of two qutrits i.e. when $d=3$.  In this case $d^{\prime}=d^{\prime\prime}=\sqrt{3}$
and the set of projectors corresponding to local projective measurement
forms the four parameter family. In the explicit parametrization we  have (see e.g.
\cite{M})
\begin{equation*}
\begin{split}
&P_{1}=\begin{pmatrix}\cos^{2}\te \sin^{2}\vf&e^{-i(\psi-\chi)}a(\te,\vf)&e^{i\chi}b(\te,\vf)\\[1mm]
e^{i(\psi-\chi)}a(\te,\vf)&\sin^{2}\te \sin^{2}\vf&e^{i\psi}c(\te,\vf)\\[1mm]
e^{-i\chi}b(\te,\vf)&e^{-i\psi}c(\te,\vf)&\cos^{2}\vf
\end{pmatrix},\\[2mm]
&P_{2}=\begin{pmatrix}\cos^{2}\te \cos^{2}\vf&e^{-i(\psi-\chi)}d(\te,\vf)&-e^{i\chi}b(\te,\vf)\\[1mm]
e^{i(\psi-\chi)}d(\te,\vf)&\sin^{2}\te \cos^{2}\vf&-e^{i\psi}c(\te,\vf)\\[1mm]
-e^{-i\chi}b(\te,\vf)&-e^{-i\psi}c(\te,\vf)&\sin^{2}\vf
\end{pmatrix},\\[2mm]
&P_{3}=\begin{pmatrix}\sin^{2}\te&-\frac{1}{2}e^{-i(\psi-\chi)}\sin 2\te&0\\[1mm]
-\frac{1}{2}e^{i(\psi-\chi)}\sin 2\te&\cos^{2}\te&0\\[1mm]
0&0&0
\end{pmatrix},
\end{split}
\end{equation*}
where
\begin{equation*}
\begin{split}
&a(\te,\vf)=\frac{1}{2}\sin 2\te \sin^{2}\vf,\\
&b(\te,\vf)=\frac{1}{2}\cos\te \sin 2\vf,\\
&c(\te,\vf)=\frac{1}{2}\sin\te \sin 2\vf,\\
&d(\te,\vf)=\frac{1}{2}\sin 2\te \cos^{2}\vf
\end{split}
\end{equation*}
and $\te,\, \vf,\, \chi\in [-\pi,\pi],\, \psi\in [-\pi/2, \pi/2]$.
\par
Our first attempt is  to compute quantum discord for states with diagonal
correlation matrix $T$ and matrix elements
\begin{equation}
t_{j}=t\,\ee_{j},\quad j=1,\ldots , 8, \label{t}
\end{equation}
where
\begin{equation*}
\ee_{j}=\W{+1}{j=1,3,4,6,8}{-1}{j=2,5,7}.
\end{equation*}
This kind of correlation matrix corresponds for example to qutrit Bell states and Werner states. Notice that
in this case
\begin{equation}
\begin{split}
&\sum\limits_{j,k}t_{j}t_{k}\ip{Me_{j}}{Me_{k}}\,\I_{3}\otimes\ip{e_{j}\ast e_{k}}{\la}\\
&=t^{2}\,\sum\limits_{j,k}\ee_{j}\ee_{k}\ip{e_{j}}{Me_{k}}\,\I_{3}\otimes \ip{e_{j}\ast e_{k}}{\la}\\
&=t^{2}\,\sum\limits_{j,k}\ip{e_{j}}{IMI\,e_{k}}\,\I_{3}\otimes\ip{e_{j}\ast e_{k}}{\la},
\end{split}\label{suma}
\end{equation}
where
\begin{equation*}
I=\mr{diag}\,(\ee_{1},\ldots, \ee_{8}).
\end{equation*}
Since
\begin{equation*}
\begin{split}
&\sum\limits_{j}\ip{e_{j}}{IMI\,e_{k}}\,\I_{3}\otimes \ip{e_{j}\ast e_{k}}{\la}\\
&=\I_{3}\otimes \ip{\sum\limits_{j}\ip{e_{j}}{IMI\,e_{k}}e_{j}\ast e_{k}}{\la}\\
&=\I_{3}\otimes \ip{IMI\,e_{k}\ast e_{k}}{\la},
\end{split}
\end{equation*}
the sum (\ref{suma}) is equal to
\begin{equation*}
t^{2}\, \I_{3}\otimes \sum\limits_{k}\ip{IMI\,e_{k}\ast e_{k}}{\la}.
\end{equation*}
By a direct computation one can check that
\begin{equation*}
\sum\limits_{k}IMI\,e_{k}\ast e_{k}=0,
\end{equation*}
so also the sum (\ref{suma}) is equal to zero. Moreover, since
\begin{equation*}
\sum\limits_{j}\, M\,e_{j}\ast M\,e_{j}=0,
\end{equation*}
we have
\begin{equation*}
\sum\limits_{j}t_{j}^{2}\,\ip{M\,e_{j}\ast M\,e_{j}}{\la}\otimes \I_{3}=0.
\end{equation*}
Notice that
\begin{equation*}
\sum\limits_{j=1}^{8}\ip{Me_{j}}{Me_{j}}=\sum\limits_{j=1}^{8}\ip{Me_{j}}{e_{j}}=\tr\, M,
\end{equation*}
so
\begin{equation*}
\begin{split}
Q(M)=\frac{1}{81}\,\bigg[&\frac{4}{9}t^{2}\,\tr\, M\,\I_{3}\otimes \I_{3}\\
&+\frac{1}{3}\,\sum\limits_{j,k=1}^{8}t_{j}t_{k}\,\ip{Me_{j}\ast Me_{k}}{\la}\otimes \ip{e_{j}\ast e_{k}}{\la}\\
&-\frac{1}{3}\,\sum\limits_{j,k=1}^{8}t_{j}t_{k}\,\ip{Me_{j}\wedge Me_{k}}{\la}\otimes \ip{e_{j}\wedge e_{k}}{\la}\,\bigg]
\end{split}
\end{equation*}
and
\begin{equation*}
\tr\, Q(M)=\frac{4t^{2}}{9\cdot 81}\,\tr\, M\cdot\tr \I_{3}\cdot \tr \I_{3}
=\frac{4t^{2}}{81}\,\tr M.
\end{equation*}
Since $M=\I_{8}-A$ projects on six-dimensional subspace of $\R^{8}$ \cite{ZCL}, $\tr M =6$ and
\begin{equation*}
\tr Q(M)=\left(\frac{2}{3}\right)^{3}\, t^{2}.
\end{equation*}
By the similar, but more involved computations, one can  check that
\begin{equation*}
\tr\, Q(M)^{k}=q_{k}\,t^{2k},\quad k=2,\ldots, 9,
\end{equation*}
where $q_{k}$ are constants. In particular
\begin{equation*}
\tr Q(M)^{k}=\tr Q(M_{0})^{k},\quad k=1,\ldots ,9,
\end{equation*}
where $M_{0}$ denotes the matrix $M$ with all parameters equal to
zero. From that, it follows that the eigenvalues of $Q(M)$ and $Q(M_{0})$
are the same \cite{L} and the distance between $\ro$ and
$\PP_{\cA}(\ro)$ is  constant. Thus for such class of states to
compute quantum discord we need not to minimize over all matrices
$M$ and it is enough to find the trace norm of $\sqrt{Q(M_{0})}$.
Next we consider two examples of states with the correlation matrix
satisfying (\ref{t}).
\subsection{Qutrit Bell state}
We start with
the \textit{Bell state} of two qutrits i.e the maximally entangled
pure state given by the vector
\begin{equation*}
\Psi_{0}=\frac{1}{\sqrt{3}}\,\sum\limits_{k=1}^{3}\vf_{k}\otimes \vf_{k},
\end{equation*}
where $\{\vf_{k}\}$ is the standard orthonormal basis in $\C^{3}$. The correlation matrix corresponding to this state is given by
\begin{equation*}
T=\mr{diag}\,\left(\frac{3}{2},\,-\frac{3}{2},\,\frac{3}{2},\,\frac{3}{2},\,-\frac{3}{2},\,\frac{3}{2},\,-\frac{3}{2},\,\frac{3}{2}\right).
\end{equation*}
One can check that in this case
\begin{equation}
Q(M_{0})=\frac{1}{9}\,\begin{pmatrix}2&0&0&0&1&0&0&0&1\\0&0&0&0&0&0&0&0&0\\0&0&0&0&0&0&0&0&0\\0&0&0&0&0&0&0&0&0\\
1&0&0&0&2&0&0&0&1\\0&0&0&0&0&0&0&0&0\\0&0&0&0&0&0&0&0&0\\0&0&0&0&0&0&0&0&0\\1&0&0&0&1&0&0&0&2
\end{pmatrix}\label{QBell}
\end{equation}
and
\begin{equation*}
\mr{sp}\, Q(M_{0})=\big\{\frac{4}{9},\, \frac{1}{9},\, \frac{1}{9},0,0,0,0,0,0\big\},
\end{equation*}
so
\begin{equation*}
\fala{D}_{1}({\Psi_0})=\tr\,\sqrt{Q(M_{0})}=\frac{4}{3}.
\end{equation*}
It is natural to demand  that the quantum discord of any maximally entangled state should be equal to $1$, so we introduce
\textit{normalized geometric measure of qutrit discord} $D_{1}(\ro)$, defined as
\begin{equation*}
D_{1}(\ro)=\frac{3}{4}\,\fala{D}_{1}(\ro).
\end{equation*}
Obviously  $D_{1}(\Psi_{0})=1$.
\subsection{Qutrit Werner states}
 As a second example we shall consider the family of qutrit \textit{Werner states}
\begin{equation}
\ro_{W}=(1-p)\,\frac{\I_{9}}{9}+
p\,\ket{\Psi_{0}}\bra{\Psi_{0}},\quad p\in [0,1].\label{W}
\end{equation}
The states (\ref{W}) have interesting properties. For $p\leq 1/4$,
$\ro_{W}$ are PPT states, whereas for $p>1/4$ they are  NPPT. In
fact such Werner states are distillable, since they violate reduction
criterion of separability \cite{HH}.
\par
One can check that in this case
\begin{equation*}
T=\mr{diag}\,\left(\frac{3}{2}p,\,-\frac{3}{2}p,\,\frac{3}{2}p,\,\frac{3}{2}p,\,-\frac{3}{2}p,\,\frac{3}{2}p,\,-\frac{3}{2}p,\,\frac{3}{2}p\right)
\end{equation*}
so the corresponding matrix $Q(M_{0})$ is just the matrix (\ref{QBell}) multiplied by the factor $p$ and
\begin{equation*}
D_{1}(\ro_{W})=p.
\end{equation*}
\par
It is instructive to compare just obtained measure of quantum discord with other measures of quantum correlations. First consider
Hilbert -Schmidt norm geometric discord $D_{2}(\ro)$, which in the case of two qutrits is defined as (see e.g \cite{GA})
\begin{equation*}
D_{2}(\ro)=\min\limits_{\PP_{\cA}}\, \frac{3}{2}\,||\ro-\PP_{\cA}(\ro)||_{2}^{2},
\end{equation*}
where
\begin{equation*}
||\sigma||_{2}^{2}=\tr\,(\sigma \sigma^{\ast}).
\end{equation*}
 For the states considered in this subsection
\begin{equation*}
D_{2}(\ro)=\frac{3}{2}\,\tr\, Q(M)=\frac{4}{9}\,t^{2}.
\end{equation*}
In particular, for the Werner state
\begin{equation}
D_{2}(\ro_{W})=p^{2},\label{D2W}
\end{equation}
and
\begin{equation*}
D_{1}(\ro_{W})=\sqrt{D_{2}(\ro_{W})}.
\end{equation*}
The result (\ref{D2W})  was previously obtained
in \cite{Y}, where the authors used minimization over all local measurements, which as we have shown, is not needed.
\par
Now we discuss the relation between $D_{1}$ and the measure of entanglement given by negativity, which in the case of
two-qutrits is defined as \cite{VW}
\begin{equation*}
N(\ro)=\frac{1}{2}\,(||\ro^{PT}||_{1}-1),
\end{equation*}
where $\ro^{PT}$ denotes partial transposition of the state $\ro$. If $N(\ro)>0$ then the state $\ro$ is non separable,
but negativity cannot detect bound entangled states.
For the Werner state we have
\begin{equation*}
N(\ro_{W})=\W{0}{p\leq\frac{1}{4}}{\frac{1}{3}(4p-1)}{p>\frac{1}{4}}.
\end{equation*}
Obviously
\begin{equation*}
D_{1}(\ro_{W})\geq N(\ro_{W})
\end{equation*}
which is in accordance with the general result proved in \cite{DMV}.
\section{Other examples}
\subsection{Some states with diagonal correlation matrix} Now we
consider the family of states with more general diagonal matrix $T$, not satisfying the condition (\ref{t}). Let
\begin{equation}
\ro=\begin{pmatrix}\frac{1}{3}&0&0&0&a&0&0&0&0\\[1mm]
0&0&0&0&0&0&0&0&0\\[1mm]
0&0&0&0&0&0&0&0&0\\[1mm]
0&0&0&0&0&0&0&0&0\\[1mm]
a&0&0&0&\frac{1}{3}&0&0&0&c\\[1mm]
0&0&0&0&0&0&0&0&0\\[1mm]
0&0&0&0&0&0&0&0&0\\[1mm]
0&0&0&0&0&0&0&0&0\\[1mm]
0&0&0&0&c&0&0&0&\frac{1}{3}
\end{pmatrix},\label{roac}
\end{equation}
where $a\geq 0,\, c\geq 0$. The matrix (\ref{roac}) is positive-definite if and only if $a^{2}+c^{2}\leq 1/9$,
so in polar coordinates we have
\begin{equation*}
a=r\,\cos \vartheta,\; c=r\,\sin\vartheta,\quad r\in [0,1/3],\, \vartheta\in [0,\pi/2].
\end{equation*}
The corresponding correlation matrix is given by
\begin{equation*}
T=\mr{diag}\, \left(\frac{9}{2}a,-\frac{9}{2}a,\frac{3}{2},0,0,\frac{9}{2}c,-\frac{9}{2}c,\frac{3}{2}\right).
\end{equation*}
In this case  the distance between
$\ro$ and $\PP_{\cA}(\ro)$ is not constant and to compute
$D_{1}(\ro)$ we must minimize $\tr\,\sqrt{Q(M)}$ over all projectors
$M$. However numerical computations show that the minimum is
achieved for $M=M_{0}$.  Since
\begin{equation*}
Q(M_{0})=\begin{pmatrix}a^{2}&0&0&0&0&0&0&0&ac\\
0&0&0&0&0&0&0&0&0\\
0&0&0&0&0&0&0&0&0\\
0&0&0&0&0&0&0&0&0\\
0&0&0&0&a^{2}+c^{2}&0&0&0&0\\
0&0&0&0&0&0&0&0&0\\
0&0&0&0&0&0&0&0&0\\
0&0&0&0&0&0&0&0&0\\
ac&0&0&0&0&0&0&0&c^{2}
\end{pmatrix},
\end{equation*}
and
\begin{equation*}
\mr{sp}\, Q(M_{0})=\{ a^{2}+c^{2},\, a^{2}+c^{2},\, 0,0,0,0,0,0,0\},
\end{equation*}
we have
\begin{equation*}
D_{1}(\ro)=\frac{3}{2}\,\sqrt{a^{2}+c^{2}}=\frac{3}{2}\, r.
\end{equation*}
On the other hand
\begin{equation*}
N(\ro)=a+c=r\,(\cos \vartheta+\sin\vartheta)
\end{equation*}
and obviously
\begin{equation*}
D_{1}(\ro)>N(\ro).
\end{equation*}
\subsection{States with non-diagonal correlation matrix: bound entangled states}
 Let us finally consider  the following family of states \cite{HHH}
\begin{equation}
\ro_{\al}=\frac{2}{7}\,\ket{\Psi_{0}}\bra{\Psi_{0}}+\frac{\al}{7}\,\ro_{+}+\frac{5-\al}{7}\,\ro_{-},
\label{bent}
\end{equation}
where
\begin{equation*}
\begin{split}
&\ro_{+}=\frac{1}{3}\,\left[P_{\vf_{1}\otimes \,\vf_{2}}+P_{\vf_{2}\otimes \,\vf_{3}}+P_{\vf_{3}\otimes \,\vf_{1}}\right]\\
&\ro_{-}=\frac{1}{3}\,\left[P_{\vf_{2}\otimes \,\vf_{1}}+P_{\vf_{3}\otimes \,\vf_{2}}+P_{\vf_{1}\otimes \,\vf_{3}}\right]
\end{split}
\end{equation*}
and $0\leq \al\leq 5$. It is known that the states (\ref{bent}) are separable for $2\leq \al\leq 3$, bound entangled for $3<\la \leq 4$ and free entangled
for $4<\al\leq 5$. One can check that the marginals of $\ro_{\al}$ are maximally mixed, but the correlation matrix $T$ is not diagonal.
In fact $T$ equals to
\begin{equation*}
\frac{1}{7}\, \begin{pmatrix}3&0&0&0&0&0&0&0\\[1mm]
0&-3&0&0&0&0&0&0\\[1mm]
0&0&-\frac{3}{4}&0&0&0&0&\frac{3\sqrt{3}}{4}(2\al-5)\\[1mm]
0&0&0&3&0&0&0&0\\[1mm]
0&0&0&0&-3&0&0&0\\[1mm]
0&0&0&0&0&3&0&0\\[1mm]
0&0&0&0&0&0&-3&0\\[1mm]
0&0&-\frac{3\sqrt{3}}{4}(2\al-5)&0&0&0&0&-\frac{3}{4}
\end{pmatrix}.
\end{equation*}
In this case we have to use directly the formula (\ref{dif}). As in the previous example, numerical computations show that it is enough
to consider $Q(M_{0})$, which is equal to
\begin{equation}
Q(M_{0})=\frac{4}{441}\,\begin{pmatrix}2&0&0&0&1&0&0&0&1\\
0&0&0&0&0&0&0&0&0\\
0&0&0&0&0&0&0&0&0\\
0&0&0&0&0&0&0&0&0\\
1&0&0&0&2&0&0&0&1\\
0&0&0&0&0&0&0&0&0\\
0&0&0&0&0&0&0&0&0\\
0&0&0&0&0&0&0&0&0\\
1&0&0&0&1&0&0&0&2\end{pmatrix}.
\end{equation}
So we have
\begin{equation}
D_{1}(\ro_{\al})=\frac{3}{4}\,\tr\sqrt{Q(M_{0})}=\frac{2}{7}
\end{equation}
and we see that quantum discord does not discriminates between separable, bound entangled and free entangled
states. On the other hand  $D_{1}(\ro_{\al})>N(\ro_{\al})$, where
\begin{equation*}
N(\ro_{\al})=\W{\frac{\DS 1}{\DS 14}\, (G(\al)-5)}{\al\in [0,1]\cup
[4,5]}{\:0}{\al\in (1,4)},
\end{equation*}
with
\begin{equation*}
G(\al)=\sqrt{41-20\al+4\al^{2}}.
\end{equation*}
We can also simply compute Hilbert-Schmidt quantum discord. It is equal to
\begin{equation*}
D_{2}(\ro_{\al})=\frac{3}{2}\,\tr\, Q(M_{0})=\frac{4}{49},
\end{equation*}
so
\begin{equation*}
D_{1}(\ro_{\al})=\sqrt{D_{2}(\ro_{\al})}.
\end{equation*}
To the authors best knowledge, the above results are the first exact results giving quantum discord of
bound entangled states. The earlier known result concerns Hilbert-Schmidt distance
quantum discord and provides only the lower and upper bounds for
$D_{2}(\ro_{\al})$ \cite{RP}. In particular it was shown that
\begin{equation}
D_{2}(\ro_{\al})\geq \W{\:\frac{\DS 4}{\DS 49}}{\al\in
[0,\al_{-}]\cup [\al_{+},5]}{\frac{\DS 1}{\DS
49}\,(9-5\al+\al^{2})}{\al\in (\al_{-},\al_{+})}\label{lb}
\end{equation}
and  the bound (\ref{lb}) is consistent  with the obtained value of $D_{1}(\ro_{\al})$.
\par
The family of states (\ref{bent}) is interesting also for another reason. When we have non-diagonal
correlation matrix $T$, we can always apply to it singular value decomposition
\begin{equation*}
T=V\,T_{0}\,W,
\end{equation*}
where $V,\, W$ are orthogonal matrices and
\begin{equation*}
T_{0}=\mr{diag}\,(s_{1},s_{2},\ldots,s_{d^{2}-1}),
\end{equation*}
with matrix elements $s_{k}$ given by singular values of $T$. In the case of qubits ($d=2$), this procedure always leads
to locally equivalent states, so we can restrict the analysis to the case of diagonal correlation matrix.
For qudits it is generally not true and the states (\ref{bent}) are the explicit counterexamples. To show this, we notice that
the singular values of the correlation matrix of the states (\ref{bent}) are given by
\begin{equation}
s_{1}=\cdots = s_{6}=\frac{3}{7},\, s_{7}=s_{8}=\frac{3}{28}\sqrt{1+3(2\al-5)^{2}}.\label{sv}
\end{equation}
Then we take $T_{0}$  defined by the sequence (\ref{sv}) and try to construct a
state using the formula (\ref{9state}), but we end with the matrix which not positive-definite. Thus there is no
equivalent description of the family (\ref{bent}) by the states with diagonal correlation matrices.
\section{Conclusions}
We have studied  behaviour of the geometric discord based on the trace norm
in the system of two qutrits. Analysis of such a system is the first non-trivial step
in extending the two-qubit theory of quantum correlations to the
general case of  $d$-level systems. We  have computed geometric discord for some  interesting
families of two-qutrit states, such as maximally entangled Bell
states, Werner states and bound entangled states. Our analysis of
qutrit systems in which entanglement can be bound or free, show,
even more clearly then in the qubit case, that  discord and
entanglement describe  different aspects of quantum correlations in
composed systems.

\end{document}